\documentclass[
twocolumn,
showpacs,preprintnumbers,
superscriptaddress,
showkeys,
amsmath,amssymb,
aps,prl
]{revtex4-1}

\usepackage{graphicx}
\usepackage{dcolumn}
\usepackage{bm}
\usepackage{hyperref}
\usepackage{color,soul}
\usepackage{amsmath}

\begin{document}


\title{Spin accumulation and dynamics in inversion-symmetric van der Waals crystals}

\author{M. H. D. Guimar\~aes}
 \email{m.h.d.guimaraes@tue.nl}
\affiliation{Department of Applied Physics, Eindhoven University of Technology, Eindhoven, The Netherlands}
\author{B. Koopmans}
 \email{b.koopmans@tue.nl}
\affiliation{Department of Applied Physics, Eindhoven University of Technology, Eindhoven, The Netherlands}

\date{\today}

\begin{abstract}
Inversion symmetric materials are forbidden to show an overall spin texture in their band structure in the presence of time-reversal symmetry.
However, in van der Waals materials which lack inversion symmetry within a single layer, it has been proposed that a layer-dependent spin texture can arise leading to a coupled spin-layer degree of freedom.
Here we use time-resolved Kerr rotation in inversion symmetric WSe$_{2}$ and MoSe$_{2}$ bulk crystals to study this spin-layer polarization and unveil its dynamics.
Our measurements show that the spin-layer relaxation time in WSe$_2$ is limited by phonon-scattering at high temperatures and that the inter-layer hopping can be tunned by a small in-plane magnetic field at low temperatures, enhancing the relaxation rates.
We find a significantly lower lifetime for MoSe$_{2}$ which agrees with theoretical expectations of a spin-layer polarization stabilized by the larger spin-orbit coupling in WSe$_2$.
\end{abstract}

\maketitle


Crystal symmetries can define many material properties.
The combination of spatial inversion and time-reversal symmetries implies a double spin degeneracy of the electronic states for a certain $k$ vector.
In the presence of structural or bulk inversion asymmetry and spin-orbit coupling (SOC), this degeneracy is lifted by the Rashba or Dresselhaus SOC \cite{Fabian2007}.
Recently it has been proposed that bulk inversion symmetric materials can show a spatially dependent spin texture resulting from a local inversion asymmetry\cite{Zunger2014}, which in the special case of layered transition metal dichalcogenides (TMDs) results in coupled spin, layer, and valley degrees of freedom \cite{Gong2013}.
Studies using photoemission spectroscopy showed spin-polarized bands \cite{King2014} and ultrafast decay (less than 1 ps) of the spin accumulation at the surface of inversion symmetric TMD crystals \cite{Wallauer2016,Hein2016,Bertoni2016,Razzoli2017}.
The high-energy beams used in these studies have a low penetration depth therefore allowing to isolate the surface signals from the bulk.
However, inversion symmetry is explicitly broken at the crystal surface which could lead to different spin texture and dynamics compared to the bulk.
Also, the crystal surface is subject to environmental conditions possibly increasing the spin-layer scattering rates.

\begin{figure*}[t]
	\includegraphics{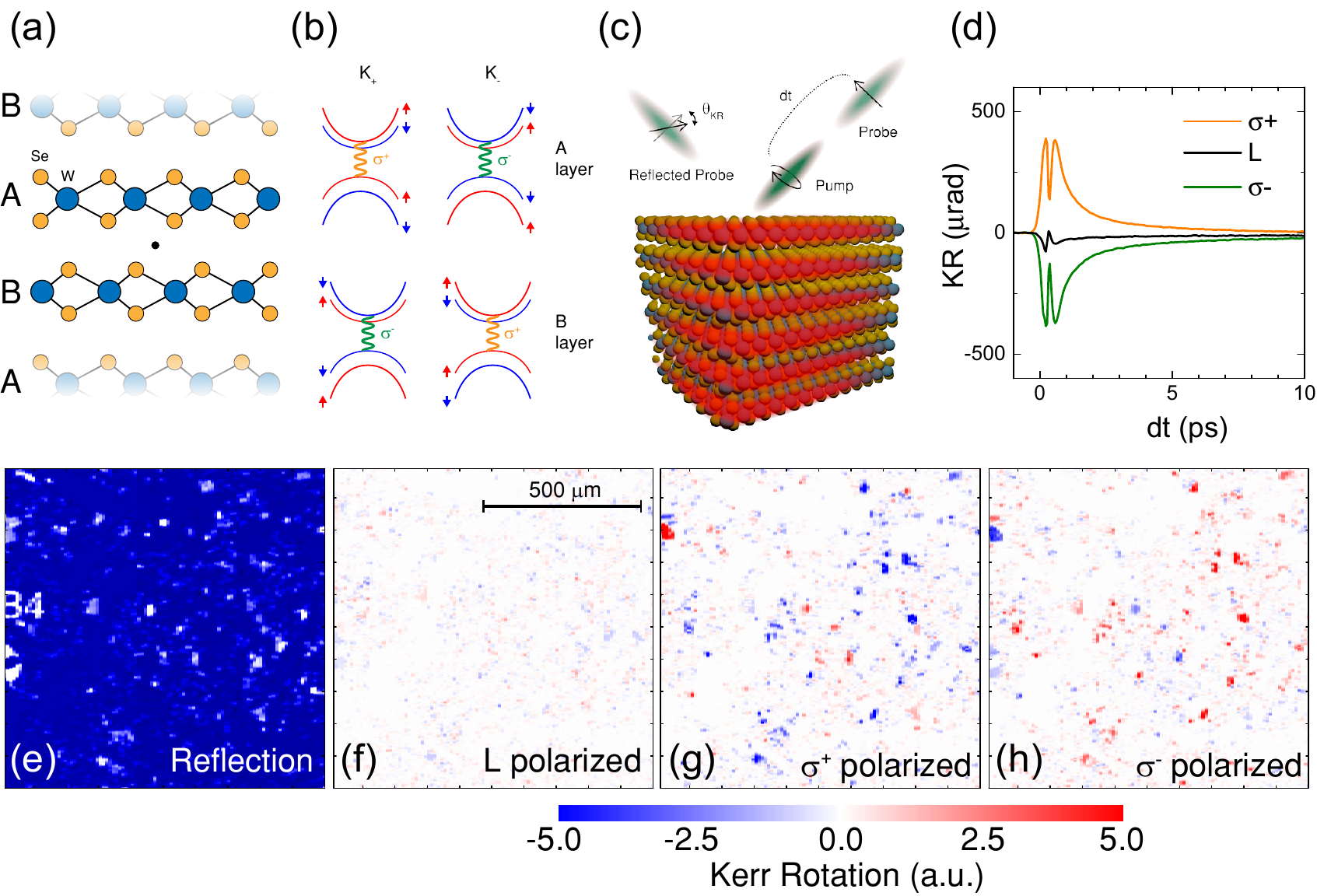}
	\caption{(a) AB-stacking of the layers of a WSe$_2$ crystal showing the inversion symmetry point. (b) Optical excitation for circularly polarized light for the A and B layers. (c) Schematics of the TRKR on a bulk TMD crystal. Depicting the rotation of the polarization ($\theta_{KR}$) of a linearly polarized probe pulse after a delay time $dt$. (d) TRKR traces for different incident pump polarizations: linearly polarized ($L$), and right-handed ($\sigma^{+}$) and left-handed ($\sigma^{+}$) circularly polarized at room temperature with $\lambda$ = 780 nm. (e) Reflection image of the sample showing several WSe$_2$ flakes of various thicknesses. The TRKR signal for $dt$ = 1 ps and $\lambda$ = 780 nm for the same region with different pump polarizations: (f) $L$, (g) $\sigma^{+}$ and (h) $\sigma^{-}$.}
	\label{fig:figure01}
\end{figure*}

Among the most studied TMDs are the semiconducting WSe$_2$ and MoSe$_2$, which have a hexagonal crystal structure with AB (Bernal) stacking where one layer (A) is rotated by 180$^\circ$ with relation to its adjacent layers (B) as shown in Fig. \ref{fig:figure01}a.
These crystals are centrosymmetric in their bulk form, but inversion asymmetric within the same layer, with the inversion operation taking one layer (and valley) into another (Fig. \ref{fig:figure01}a), therefore coupling the layer, spin, and valley degrees of freedom \cite{Gong2013,JaquePRB2014}.
TMDs are also fascinating materials for spin-logic devices due to their large SOC, spin-valley locking and optically addressable states\cite{Xiao2012, Liu2013, Xu2014}.
This leads to an interesting optically-accessible degree of freedom, where circularly polarized light couples to the same spin-polarized bands but at different valleys for different layers (Fig. \ref{fig:figure01}b)\cite{Gong2013}.

Here we show experimental evidence that the coupled layer-spin-valley degrees of freedom allow for the optical generation of spin accumulation and study its dynamics in bulk TMD crystals using time-resolved Kerr rotation (TRKR).
The long penetration depth ($>$ 50 nm) of our laser allows us to excite and probe the spin dynamics deep within the crystal and not only at its surface.
The spin dynamics in the bulk crystal is expected to be dominated by the material itself, opposed to surface and substrate-induced scattering, since the surface layers screen most environmental scattering mechanisms.
This can help to elucidate the wide range of reported spin lifetimes in atomically thin TMDs, ranging from a few ps to several ns at low temperatures \cite{Zeng2012,Mak2012,Wu2013,ZhuPNAS2014,Jones2014,Lagarde2014,Zhu2014,Mai2014,Liu2015,Yan2015,Yang2015,Hsu2015,Smolenski2016,Song2016,Plechinger2017,Yan2017,Schwemmer2017,Volmer2017,McCormick2017,Ersfeld2017}.
We show that the helicity of a pump beam determines the orientation of the spin polarization, and that the spin accumulation created by the pump beam reaches a lifetime ($\tau$) of 50 ps at low temperatures in bulk WSe$_2$.
We get further insights on the limiting factors for the relaxation mechanisms in our samples by temperature dependent measurements and applying an in-plane magnetic field, which suggests that $\tau$ is limited by phonon-scattering at temperatures above 100 K and inter-layer scattering for lower temperatures.
Finally, in order to investigate the impact of SOC on our measurements we compare WSe$_2$ and MoSe$_2$ and find a drastic difference for their respective $\tau$, with MoSe$_2$ showing much lower values, in agreement with theoretical expectations \cite{Gong2013,Liu2015}.
The demonstration of optical generation and dynamics of spin accumulation using the layer-dependent spin texture in bulk TMDs broadens the application prospects of TMD-based spintronic devices beyond the use of atomically thin TMDs, which relaxes constraints in growth and device fabrication techniques.

Our samples are fabricated using mechanical exfoliation of WSe$_2$ or MoSe$_2$ crystals (HQ graphene) onto a Si/SiO2 (100 nm) wafer.
The flakes are identified using optical microscopy and pre-defined markers on the substrate.
By shining a circularly polarized pump beam we excite spin polarized carriers with the same spin orientation but at opposite valleys for adjacent layers (Fig. \ref{fig:figure01}b).
The spin accumulation is then detected after a delay time $dt$ by using a weaker linearly polarized laser pulse (probe; Fig. \ref{fig:figure01}c) with the same wavelength as the pump ($\lambda$).
Due to the Kerr effect, the reflected probe beam shows a small ellipticity and its polarization axis is slightly rotated by an angle $\theta_{KR}$ proportional to the spin accumulation.
The sign and amplitude of the signal is determined by the polarization of the pump beam (Fig. \ref{fig:figure01}d) \cite{supinfo}.
Scanning the pump and probe beams together at a fixed $dt$ allows us to locate and determine the TRKR signal of several individual flakes on our substrate (Fig. \ref{fig:figure01}e-h).
In order to avoid excessive heating and high excitonic density effects, we use pump fluences below 10 $\mu$J/cm$^2$ \cite{Mouri2014, supinfo}.
To exclude expurious signals we subtract the TRKR traces for different pump helicities: right-handed ($\sigma^{+}$) and left-handed ($\sigma^{-}$) circularly polarized: KR$_{\sigma +}$ and KR$_{\sigma -}$, respectively to obtain the Kerr rotation induced by the optically induced spin accumulation, which we refer as KR Polarization.
The experimental setup and sample characterization are detailed in the Supplementary Information \cite{supinfo}.

\begin{figure}[h]
	\centering
		\includegraphics{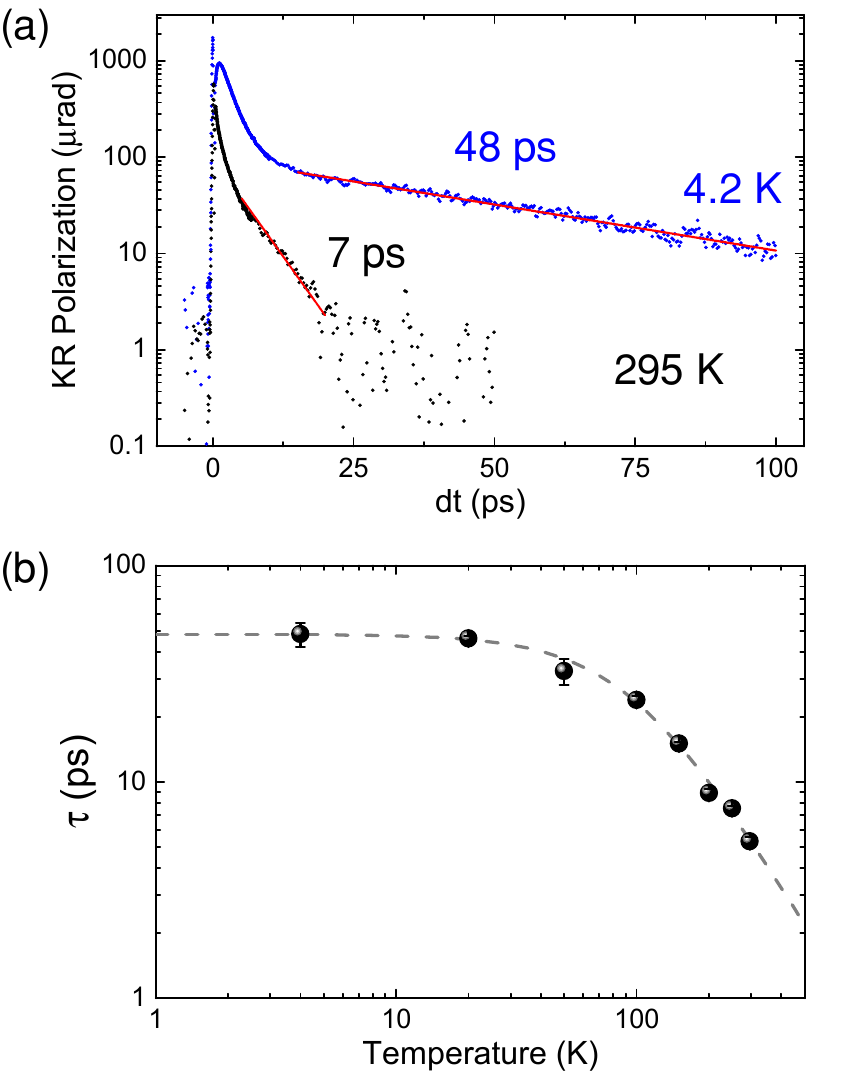}
	\caption{(a) Polarization of the Kerr rotation at 295 K (black) and 4.2 K (blue) for WSe$_2$. The red lines show the linear fits for which the decay constants ($\tau$) are extracted. (b) $\tau$ \textit{versus} $T$. The dashed gray line is a fit as explained in the main text.}
	\label{fig:figure02}
\end{figure}

The lower panels of Figure \ref{fig:figure01} show the probe reflection intensity (e) and the TRKR signal ($dt$ = 1 ps) for different polarizations of the pump beam: linearly polarized ($L$) (f), $\sigma^{+}$ (g) and $\sigma^{-}$ (h) for the same 1x1 mm$^2$ region of exfoliated WSe$_2$ flakes.
While the Kerr rotation is negligible for most flakes with a linearly polarized pump, clear signals can be observed for when the pump beam is circularly polarized.
The TRKR signal continuously changes with pump polarization (see Supplementary Information \cite{supinfo}) and upon reversal of the pump helicity the TRKR signals reverse sign, confirming that the induced spin polarization reverses sign with pump helicity.
It is interesting to note that different flakes show signals with signs for a fixed pump polarization which could be explained by a small shift in the A exciton peak for different flakes due to, for example, doping \cite{supinfo}.

Due to the high SOC in WSe$_2$, an increased momentum scattering should result in a shorter spin lifetime.
Phonon-scattering has been shown to be a limiting factor for the electronic mobility in TMDs, therefore it is also expected it will limit the relaxation time in our measurements.
This is indeed what we observe.
At room temperature we measure two decay constants in the TRKR signal, of approximately 2 and 5 ps and, as the temperature is reduced, we observe an increase on both the TRKR signal amplitude and the long time-constant ($\tau$), to about 50 ps at 4.2 K (Fig. \ref{fig:figure02}a), while the fast component does not show any systematic change \cite{supinfo}.
For these measurements $\lambda$ was kept at the A-exciton resonance as a function of $T$, but we find that the overall behavior does not depend on the specific laser energy around the A-exciton resonance (see Supplementary Information for details\cite{supinfo}).
The measurements shown here were obtained on a representative WSe$_2$ flake ($\approx$75 nm thick). However, all other flakes studied show similar results.
Multiple components TRKR signals were also obtained in monolayer WSe$_2$ \cite{Zhu2014, Song2016, Yan2017, Plechinger2017} where the fast decays have been attributed to exciton recombination or electron-hole exchange interactions.
Here we focus on the slow component of the TRKR signal unless specified.

We find that the $T$ dependence of $\tau$ is well described by: $\tau^{-1} = \tau_{0}^{-1} + \tau_{T}^{-1}$, where $\tau_{0}$ is a temperature-independent decay and $\tau_{T} = C \times T^{-\alpha}$ where $C$ and $\alpha$ are positive constants.
From fitting our experimental data (Fig. \ref{fig:figure02}b) we obtain $\tau_{0} = (48 \pm 2)$ ps, and $\alpha = (1.8 \pm 0.1)$, where a value of $\alpha \approx$ 1.5 implies isotropic lattice scattering by optical phonons \cite{Fivaz1967}.
Interestingly, while the electronic mobility in n-type bulk WSe$_2$ crystals showed exponents of about 2.5 \cite{Fivaz1967}, $\alpha \approx$ 2.0 (1.0) has been shown for n-type (p-type) WSe$_2$ monolayers \cite{Xu2016}, pointing to different scattering mechanisms for charge and spin carriers.
The energy of the dominant scattering phonon ($E_{\text{ph}}$) can be estimated by assuming scattering by phonon absorption, resulting in $E_{\text{ph}} = (21 \pm 3)$ meV \cite{supinfo}, which is approximately the energy for the nearly degenerate $E_{\text{2g}}$ and $A_{\text{1g}}$ optical phonon modes (30 meV).
This value is similar to the one obtained in monolayer WSe$_2$ \cite{Hsu2015}, indicating the similar dominant scattering phonon in the bulk crystal.
%

\begin{figure}[t!]
	\includegraphics{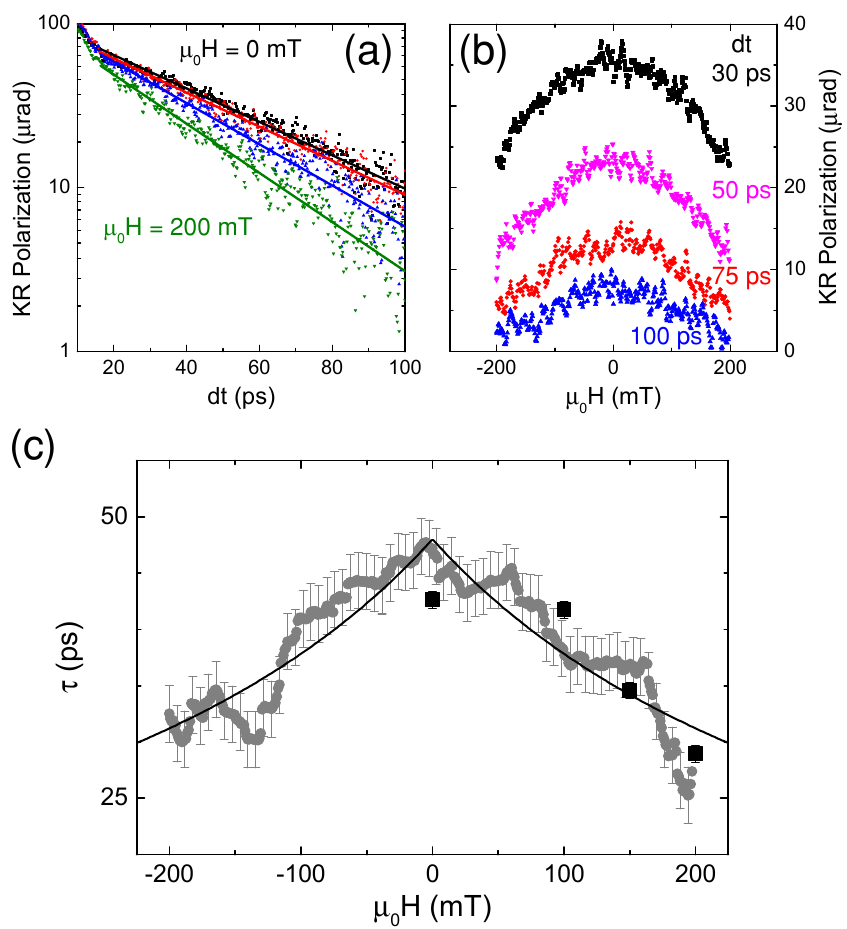}
	\caption{(a) KR Polarization as a function of $dt$ for $\mu_{0}H_{y}$ = 0 (black), 100 (red), 150 (blue), and 200 mT (green). The points are the experimental data and the solid lines are the linear fits to obtain $\tau$. (b) KR Polarization \textit{versus} $\mu_{0}H_{y}$ for $dt$ = 30 (black), 50 (magenta), 75 (red), and 100 ps (blue). (c) $\tau$ \textit{versus} $\mu_{0}H_{y}$ extracted by the measurements in (a) (black squares) and (b) (gray circles). The solid line is a theoretical prediction based on the cyclotron motion as described in the main text. The measurements were performed at 4.2 K.}
	\label{fig:figure03}
\end{figure}

The saturation of the relaxation rate at low temperatures indicates that other mechanisms than phonon scattering limit it.
The presence of in-plane magnetic field ($\mu_0 H_{y}$), in addition to increasing the mixing between the spin-up and spin-down levels, causes the charge carriers to precess in a cyclotron motion around the field enhancing inter-layer scattering \cite{supinfo}.
To explore this possibility we obtain measurements of the TRKR signal at 4.2 K as a function of $dt$ for different values of $\mu_0 H_{y}$ (Fig. \ref{fig:figure03}a), and as a function of $\mu_0 H_{y}$ for fixed values of $dt$ (Fig. \ref{fig:figure03}b)
Both measurements show a consistent decrease of $\tau$ with an increase of $\mu_0 H_{y}$, from about 50 ps at $\mu_0 H_{y}$ = 0 mT to about 30 ps for $| \mu_0 H_{y} |$ = 200 mT (Fig. \ref{fig:figure03}c).

Different studies in the literature show competing results on the effect of an in-plane magnetic field on the spin relaxation in monolayer TMDs.
While a few studies show that a small magnetic field has no effect on the total spin-valley accumulation\cite{Zeng2012,Song2016,Schwemmer2017,McCormick2017} as expected from the large spin-orbit fields ($B_{SO}$) in TMDs, other works show an increase\cite{Smolenski2016} or decrease on the spin signal and even coherent precession in their signals\cite{Yang2015,Volmer2017}, which has been attributed to spins in localized defects.
It has been proposed that the fast decay of TRKR signals with $\mu_0 H_{y}$ in TMD monolayers could be due to an enhanced inter-valley scattering \cite{Yang2015}.
Using a simple model we find that we can only explain the decay seen in our signals by this process if $\mu_0 H_{y} \approx B_{SO}$, which seems unreasonable in our case.
However, our data can be explained if we take into account that the cyclotron motion around the magnetic field induces a momentum in the z direction, forbidden in the monolayer case.
The gain in $k_{z}$ in the direction of the H point causes intermixing between different layers, enhancing this relaxation channel.
Our data follows closely the simple relation: $\tau^{-1} = \tau_{0}^{-1} + T^{-1}$ with a single adjustable parameter $\tau_{0}$ = 48 ps, and where $T = m^{*}/e| \mu_{0}H_{y} |$ is the cyclotron period and $m^{*}$ = 0.5 is the effective mass for holes in WSe$_2$ \cite{Song2013,Xu2016}.
A $1/| \mu_{0}H_{y} |$ behavior is also expected for an anisotropic g-factor $\Delta g$. Fitting our data we obtain $\Delta g = (0.65 \pm 0.1)$, larger than the factor found in monolayers ($\Delta g \approx 0.3$) \cite{Volmer2017}, indicating that the first mechanism or a combination of both are into play.
Therefore, our results imply that while phonon-scattering dominates at high temperatures, interlayer hopping plays a major role in the spin relaxation in WSe$_2$ at low temperatures.

\begin{figure}[h]
	\includegraphics{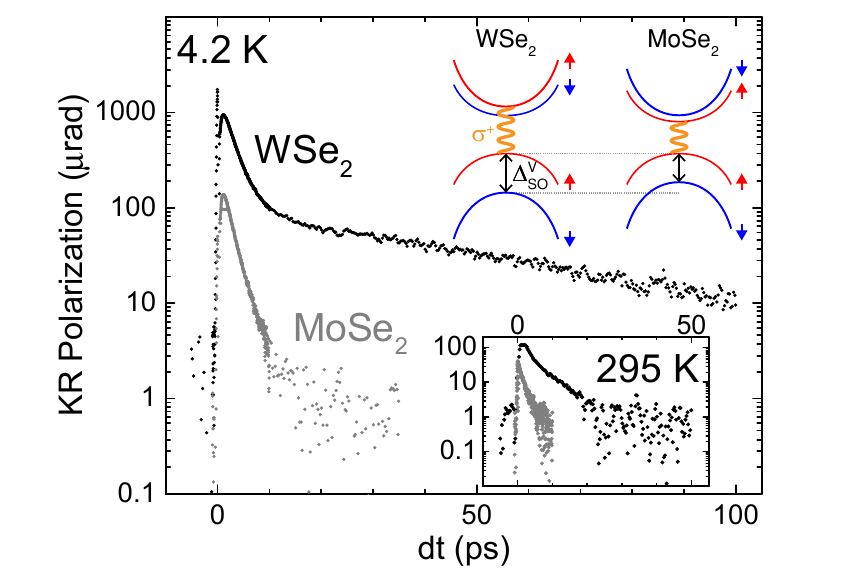}
	\caption{Kerr rotation polarization as a function of time delay for WSe$_2$ (black) and MoSe$_2$ (gray) at 4.2 K. \textit{Lower Inset:} Same as in main graph, but at room temperature (295 K) \textit{Upper Inset:} Band structure of WSe$_2$ and MoSe$_2$ near the K point showing the allowed optical transitions.}
	\label{fig:figure04}
\end{figure}

In order to understand the impact of the band structure on the spin-layer polarization we performed TRKR experiments on MoSe$_2$.
While we observe a decay time for WSe$_2$ that strongly depends on temperature, this is not the case for MoSe$_2$ (Fig. 4).
The lifetimes in MoSe$_2$ are found to be much shorter, remaining below 6 ps at low temperatures, and do not show a significant temperature dependence \cite{supinfo}.
Furthermore, while we observe two decay constants for the TRKR signal in WSe$_2$, only one is observed within our experimental limits in our MoSe$_2$ crystals.
MoSe$_2$ and WSe$_2$ are similar materials with subtle but important differences in their bandstructures.
MoSe$_2$ shows a smaller bandgap ($\Delta$ = 1.5 eV) and SOC ($\lambda_{v}$ = 180 meV and $\lambda_{c}$ = 20 meV for the valence and conduction bands, respectivelly) compared to WSe$_2$ ($\Delta$ = 1.7 eV, $\lambda_{v}$ = 450 meV and $\lambda_{c}$ = 35 meV).
Furthermore, opposite to WSe$_2$, MoSe$_2$ has its dark exciton states at higher energies than its bright exciton states (Fig. 4).

Our findings of shorter $\tau$ in MoSe$_2$ are in agreement with theoretical suggestions that the spin relaxation in inversion symmetric TMDs is suppressed by a combination of low interlayer hopping ($t_{\perp}$) and high SOC \cite{Gong2013}.
The large value of $\lambda_{v}$/$t_{\perp} \approx$ 7 in WSe$_2$ compared to $\lambda_{v}$/$t_{\perp} \approx$ 3 in MoSe$_2$ predicts a shorter lifetime for MoSe$_2$.
Moreover, it has been shown that the presence of dark excitons below the bright exciton states in WSe$_2$ helps the stabilization of the spin polarization in monolayers \cite{Baranowski2017}.
While we cannot pinpoint the importance of each mechanism in our experiments, we observe a clear difference in the TRKR signal for both materials, indicating that either (or both) mechanisms are responsible for stabilizing the spin polarization in WSe$_2$.

The observation of spin-layer polarization and its dynamics in bulk TMD crystals confirms the presence of coupled spin-layer-valley degree of freedom extending throughout the whole crystal, and not only at its surface.
Overall, our results for the spin lifetimes of several ps in bulk TMDs fall in between the several results reported in the literature for monolayers showing that the bulk TMD crystals can be seen as weakly-interacting individual monolayers.
This indicates that the use of TMDs for spintronic applications should not be limited to monolayers, but is also promising for bulk materials, relaxing several constraints in device fabrication and material growth.
The confirmation of the optically addressable layer-dependent spin texture and the possibility of manipulating the spin accumulation using a small magnetic field in bulk TMD crystals should also help the development and applications of spin-logic devices based on TMDs \cite{Zhang2016,Shao2016,MacNeill2016,Guimaraes2018}.

\begin{acknowledgments}
We thank M. L. M. Lalieu and M. J. G. Peeters, J. Francke and G. Basselmans for help with the experimental setup and technical support, and R. A. Duine and G. M. Stiehl for useful discussions and comments on the manuscript.
This work was supported by the Netherlands Organization for Scientific Research (NWO VENI 15093).
Sample fabrication was performed using NanoLab@TU/e facilities which is part of the NanoLabNL network.
\end{acknowledgments}


%

\end{document}